\documentclass[twocolumn,aps,prl,groupedaddress,showkeys]{revtex4-2}
\usepackage{bm}
\usepackage{amsfonts}
\usepackage{amsmath}
\usepackage{amssymb}
\usepackage{graphicx}
\usepackage{braket}
\usepackage{array}
\usepackage{threeparttable}
\usepackage{bigstrut}
\usepackage{lipsum}
\usepackage[colorlinks,linkcolor=blue,anchorcolor=blue,citecolor=blue,urlcolor=blue]{hyperref}
\usepackage{color}

\begin{document}

\title{Direct Extraction of Nuclear Structure Information Using Precision \\
Lithium-Ion Spectroscopy}


\author{Hua Guan$^{1,*}$}
\author{Xiao-Qiu Qi$^{2,*}$}
\author{Jian-Guo Li$^{5,8}$}
\author{Peng-Peng Zhou$^{1}$}
\author{Wei Sun$^{6,1}$}
\author{Shao-Long Chen$^{1}$}
\author{Xu-Rui Chang$^{1,8}$}
\author{Yao Huang$^{1}$}
\author{Pei-Pei Zhang$^{1}$}
\author{Zong-Chao Yan$^{3,1}$}
\author{G. W. F. Drake$^{4}$}
\author{Ai-Xi Chen$^{2}$}
\author{Zhen-Xiang Zhong$^{7,1}$}
\author{Jia-Li Wang$^{5,8}$}
\author{Nicolas Michel$^{5,8}$}
\author{Ting-Yun Shi$^{1,\dag}$}
\author{Ke-Lin Gao$^{1,\ddag}$}

\affiliation {$^1$State Key Laboratory of Magnetic Resonance and Atomic and Molecular Physics, Wuhan Institute of Physics and Mathematics, Innovation Academy for Precision Measurement Science and Technology, Chinese Academy of Sciences, Wuhan 430071, China}
\affiliation {$^2$Department of Physics, Zhejiang Sci-Tech University, Hangzhou 310018, China}
\affiliation {$^3$Department of Physics, University of New Brunswick, Fredericton, New Brunswick, Canada E3B 5A3}
\affiliation {$^4$Department of Physics, University of Windsor, Windsor, Ontario, Canada N9B 3P4}
\affiliation {$^5$Heavy Ion Science and Technology Key Laboratory, Institute of Modern Physics, Chinese Academy of Sciences, Lanzhou 730000, China.}
\affiliation {$^6$Key Laboratory of Green and High-end Utilization of Salt Lake Resources, Qinghai Institute of Salt Lakes, Chinese Academy of Sciences, Xining 810008, China}
\affiliation {$^7$Center for Theoretical Physics, School of Physics and Optoelectronic Engineering, Hainan University, Haikou 570228, China}
\affiliation {$^8$University of Chinese Academy of Sciences, Beijing 100049, China}

\begin{abstract}
Accurately describing nuclear interactions within atomic nuclei remains a challenge, which hinders our exploration of new physics beyond the Standard Model. However, these nuclear interactions can be characterized by nuclear parameters such as the Zemach radius and the electric quadrupole moment, which are reflected in atomic spectra. Our work has achieved high-precision measurements of lithium ion hyperfine splittings at the level of $10$~kHz, and directly extracted these important nuclear structure parameters. We observed significant discrepancies between our results and both nuclear theory and molecular spectra regarding the electric quadrupole moment. The result for $^7$Li deviated by $2.3\sigma$ from the currently recommended value, whereas the result for $^6$Li deviated by up to $6.2\sigma$ from the recommended value determined by molecular spectroscopy. These discrepancies motivated us to conduct independent calculations based on nuclear structure theory, which provided support for the results obtained from ion spectroscopy. Our results provide valuable information for characterizing nuclear forces, serve as sensitive benchmarks for testing nuclear structure theories, and enable critical comparisons with both electron-nuclear scattering and molecular spectroscopy.
\end{abstract}


\date{\today}
\maketitle

\emph{Introduction.}---
The atomic nucleus is formed by the combination of protons and neutrons under the nuclear force, and its internal structure and motion are primarily described by various phenomenological models. As of now, a universal nuclear structure theory has not been established due to a lack of comprehensive knowledge of nuclear forces. The Zemach radius~\cite{Zemach1956}, which describes the distribution of nuclear charge and magnetization in atomic nuclei, and the electric quadrupole moment~\cite{Stone2016}, which characterizes the shape deformation of atomic nuclei, are both external manifestations of the internal structure and motion of nuclei. Their influences are reflected in the hyperfine splittings (hfs) of atomic spectra. 
Furthermore, the Zemach radius can be used to extract the magnetic radius that characterizes the nuclear magnetic distribution inaccessible through spectroscopic methods, and this magnetic radius is more sensitive to larger momentum transfers in electron scattering~\cite{LIN2024}.
Thus, the precise determination of these fundamental nuclear properties through precision spectroscopy is of significant importance in testing nuclear models, exploring precise nuclear forces, and establishing a comprehensive nuclear structure theory~\cite{Dickopf2024,STAR2024}.

The helium-like Li$^+$ ion is considered a fundamental atomic system due to its two-electron nature, encompassing fine and hyperfine structure, which can be experimentally measured and theoretically calculated with high precision. This system serves as a platform for testing bound-state quantum electrodynamic (QED) theory~\cite{Yan1995,Yan2008,Pachucki2010,Pachucki2019,Wu2022}, determining nuclear properties~\citep{Neugart2008,Drake2013,Yang2023}, and exploring potential new physics beyond the Standard Model~\cite{Drake2021}. 
For example, the latest precision spectroscopic methods have determined the Zemach radii of $^6$Li and $^7$Li with an accuracy of 1\%~\cite{Sun2023,Pachucki2023}. However, the discrepancy between the spectral result and the nuclear physics value for $^6$Li has increased to $8\sigma$~\cite{Yerokhin2008}, presenting an anomaly that cannot be explained by Bohr-Weisskopf theory, especially considering the good agreement observed in the case of $^7$Li~\cite{Puchalski2013,Qi2020,Sun2023,Pachucki2023}.
The ultimate resolution of this discrepancy will enhance our understanding of nuclear properties, particularly in exploring the characteristics of odd-odd nuclei, and may also lead to the discovery of previously unknown weak physical effects.

Moreover, in nuclear astrophysics, the cosmological lithium problems are well-known~\cite{Boyd2010,Cyburt2016}. In detail, the observed abundance of $^7$Li and the abundance ratio of $^6$Li/$^7$Li differ significantly from the predictions of big-bang nucleosynthesis theory~\cite{Rijal2019,Ali2022}. Currently, no satisfactory explanation has been found for these discrepancies. It is important to note that the abundance of elements produced by nuclear reactions is directly influenced by the astrophysical $S$ factor, which characterizes the nuclear reaction cross section and is closely related to the electric quadrupole moment of the nucleus~\cite{Grassi2017}.
Recent measurements of the hyperfine structure spectrum of $^6$Li$^+$ ions have yielded a rough estimate of the nuclear quadrupole moment, $-0.38(20)$~fm$^2$~\cite{Sun2023}, which differs significantly from the currently recommended value $-0.0806(6)$~fm$^2$~\cite{Cederberg1998,Stone2016}.
This new value is in good agreement with $-0.35(6)$ fm$^2$ from the Green's function Monte Carlo (GFMC) calculation~\cite{Pieper2001,Pieper2002}. However, there is considerable divergence in the theoretical values, owing to a sensitive dependence on the details of effective nuclear interaction, ranging from $-0.061$ fm$^2$ \cite{Nortershauser2011} and $-0.066(40)$ fm$^2$ \cite{Forssen2009} to $-0.35(6)$~fm$^2$~\cite{Nortershauser2011}. For example, the value $-0.061$ fm$^2$ from fermionic molecular dynamics calculations \cite{Nortershauser2011} becomes positive if the tensor part of the interaction is not included. 

The discrepancies mentioned above are puzzling and motivate us to investigate the nuclear structure information of $^6$Li and $^7$Li from both precision spectroscopy and nuclear structure theory.

\begin{figure}[tbp!]
\includegraphics[scale=0.3,trim=20 40 0 60,clip]{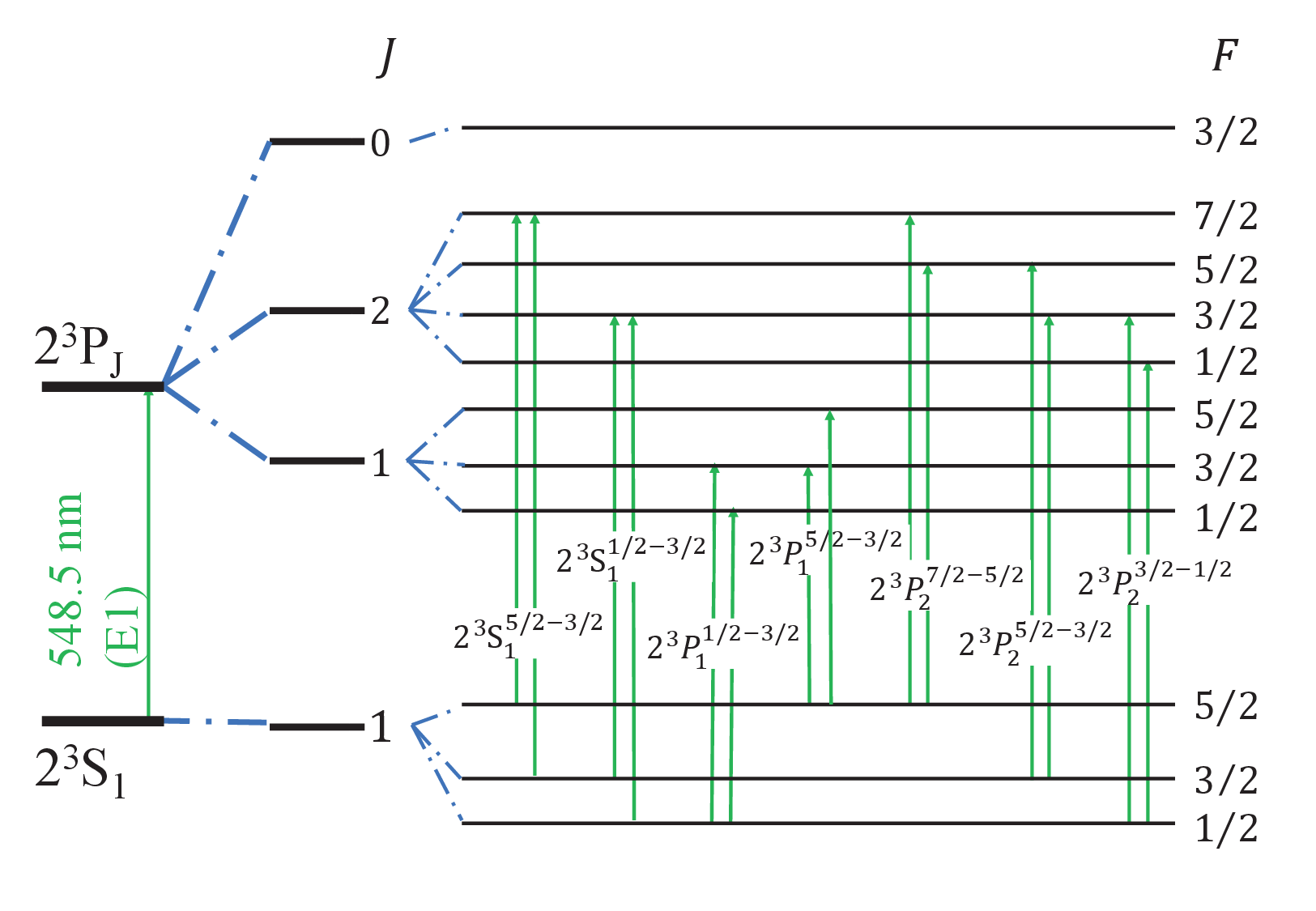}
\caption{Fine and hyperfine structures for the $2\,^3\!S_1$ and $2\,^3\!P_J$ states of $^7$Li$^+$ ion (not to scale).}
\label{fig:1}
\end{figure}


\begin{figure}[htbp!]
\centering
\begin{minipage}{0.3\textwidth}
\includegraphics[width=\textwidth]{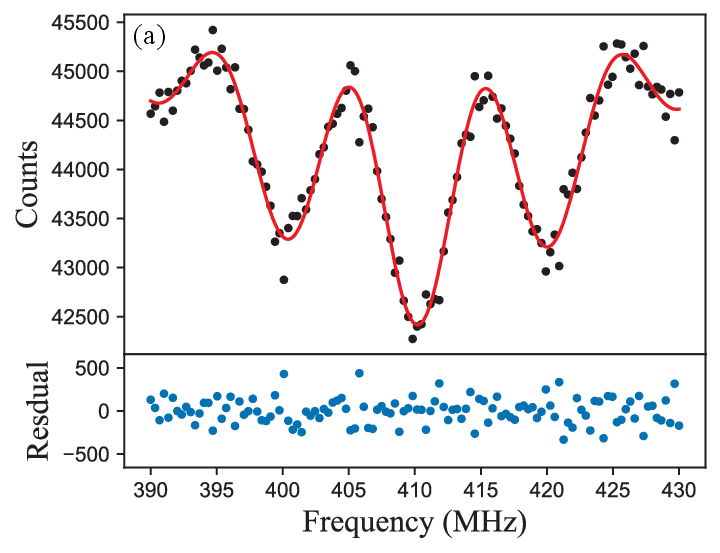}
\end{minipage}
\quad
\begin{minipage}{0.5\textwidth}
\includegraphics[width=\textwidth]{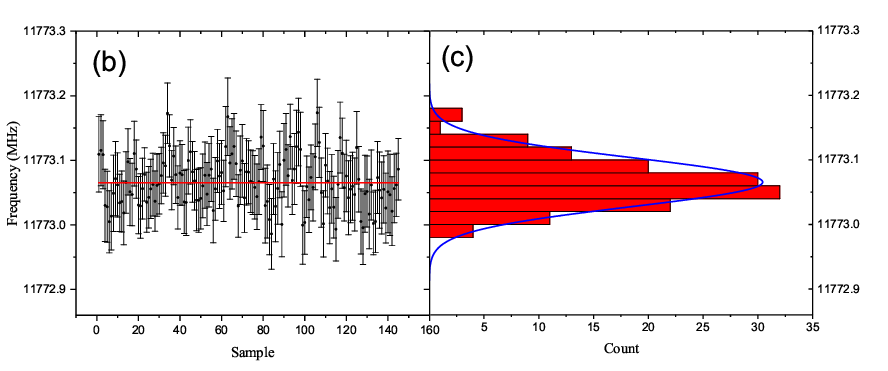}
\end{minipage}
\caption{(a). The Ramsey fringes for a single measurement. (b) and (c). The data distribution of hfs $2\,^3\!P_2^{5/2-7/2}$ and its Gaussian fitting of histograms of data distribution.}
\label{Ramsey}
\end{figure}
 
\emph{Precision Spectroscopy.}---
Hyperfine structure refers to the further splitting of atomic energy levels caused by nuclear spin. Figure~\ref{fig:1} presents a schematic diagram illustrating the energy levels of $^7$Li$^+$, where the hyperfine structure is superimposed on the fine structure. The hyperfine level splitting can be expressed as
\begin{equation}\label{eq:1}
\begin{aligned}
E_{\rm hfs} = E_{p}+E_{\rm ZM}+E_{Q_d}\,,
\end{aligned}
\end{equation}
where $E_{p}$ represents the energy of hfs obtained for the case of a point nucleus, and $E_{\rm ZM}$ and $E_{Q_d}$ are contributions generated by the nuclear Zemach radius~\cite{Zemach1956} and quadrupole moment, respectively.
Since the contribution of the nuclear quadrupole moment to the $2\,^3\!S_1$ state is negligibly small, we can first extract the Zemach radius by measuring the hyperfine structure of the $2\,^3\!S_1$ state. Subsequently, the obtained Zemach radius, along with the corresponding measurements of the hyperfine structure of the $2\,^3\!P_J$ state, will be used to determine the nuclear quadrupole moment.

\begin{table*}[htbp!]
  \caption{\label{tab:1} Experimental and theoretical hyperfine intervals in the $2\,^3\!S_1$ and $2\,^3\!P_J$ states of $^7$Li$^+$, in MHz. $\overline{\Delta}(2\,^3\!P)$ represents the root-mean-square deviation of the other results from the theoretical values of Qi {\it et al.} \cite{Qi2020}.}
   \begin{ruledtabular}
   \begin{tabular}{lccccccc}
   \multicolumn{1}{l}{}
  &\multicolumn{4}{c}{Experiment}
  &\multicolumn{2}{c}{Theory}\\
  \cline{2-5} \cline{6-7}
   \multicolumn{1}{l}{hfs interval}
  &\multicolumn{1}{c}{K{\"o}tz {\it et al.}~\cite{Kotz1981,Kowalski1983}}
  &\multicolumn{1}{c}{Clarke {\it et al.}~\cite{Clarke2003}}
  &\multicolumn{1}{c}{Guan {\it et al.}~\cite{Guan2020}}
  &\multicolumn{1}{c}{This work}
  &\multicolumn{1}{c}{Drake {\it et al.}~\cite{Riis1994}}
  &\multicolumn{1}{c}{Qi {\it et al.}~\cite{Qi2020}}\\
  \hline
  $2\,^3\!S_1^{1/2-3/2}$ &$11890.018(40)$ &$11891.22(60)$ &$11890.088(65) $ &$11890.012(15)$ &$11890.013(38)$ &                \\
  $2\,^3\!S_1^{3/2-5/2}$ &$19817.673(40)$ &$19817.90(93)$ &$19817.696(42) $ &$19817.677(12)$ &$19817.680(25)$ &                \\
  $2\,^3\!P_1^{1/2-3/2}$ &$ 4237.8(10)  $ &$ 4239.11(54)$ &$ 4238.823(111)$ &$ 4238.968(13)$ &$ 4238.86(20) $ &$ 4238.920(49)$ \\
  $2\,^3\!P_1^{3/2-5/2}$ &$ 9965.2(6)   $ &$ 9966.30(69)$ &$ 9966.655(102)$ &$ 9966.437(11)$ &$ 9966.14(13) $ &$ 9966.444(34)$ \\
  $2\,^3\!P_2^{1/2-3/2}$ &$ 6203.6(5)   $ &$ 6204.52(80)$ &$ 6203.319(67) $ &$ 6203.320(14)$ &$ 6203.27(30) $ &$ 6203.408(95)$ \\
  $2\,^3\!P_2^{3/2-5/2}$ &$ 9608.7(20)  $ &$ 9608.90(49)$ &$ 9608.220(54) $ &$ 9608.238(11)$ &$ 9608.12(15) $ &$ 9608.311(54)$ \\
  $2\,^3\!P_2^{5/2-7/2}$ &$11775.8(5)   $ &$11774.04(94)$ &$11772.965(74) $ &$11773.066(11)$ &$11773.05(18) $ &$11773.003(55)$ \\
  $\overline{\Delta}(2\,^3\!P)$ &$1.47  $ &$0.74        $ &$0.12          $ &$0.06         $ &$0.18         $ &$             $ \\
  \end{tabular}
  \end{ruledtabular}
\end{table*}

The present hfs measurements rely on the optical Ramsey technique~\cite{Baklanov1976,Borde1984,Bergquist1977} employing separated laser fields. The experimental setup is illustrated in the Supplemental Material~\cite{SM}. A more comprehensive account of the ion source and the corresponding laser systems for the separated fields can be found in Refs.~\cite{Sun2023,Chen2019,SM}. The Ramsey fringes for a single measurement are presented in Figure~\ref{Ramsey}(a). The data distribution of the hfs for $2\,^3\!P_2^{5/2-7/2}$, depicted in Figures~\ref{Ramsey}(b) and (c), conforms to a normal distribution, with a statistical standard deviation of approximately $4$~kHz. Each hfs is measured $100$-$200$ times over a period of $10$ hours, from which the statistical mean and uncertainty are calculated. These results and the evaluation of systematic errors, including Doppler shift, laser power, Zeeman effect, and quantum interference, are provided in the Supplemental Material~\cite{SM}.

\emph{Nuclear Structure Theory.}---The Gamow Shell Model (GSM), an advanced extension of the traditional shell model into the complex momentum 
$k$-plane~\cite{Michel:2002,Michel:2021,Jin:2021,PhysRevC.103.034305}, is employed in our nuclear structure calculations. Its theoretical foundation lies in the Berggren basis~\cite{Berggren:1968}, which includes bound states, resonances, and scattering states generated by a finite-depth potential.

We model the nucleus as valence nucleons interacting outside a closed $^4$He core. The coordinates of the valence nucleons are defined relative to the center of mass of the core, using the Cluster Orbital Shell Model (COSM) 
framework~\cite{Suzuki1988}. In the COSM frame, the GSM Hamiltonian comprises a one-body core-valence potential, a two-body interaction between the valence nucleons, and a two-body recoil term~\cite{Suzuki1988}.  

The core-valence potential is modeled using a Woods-Saxon (WS) potential that includes central, spin-orbit, and Coulomb components. The two-body interaction is represented by the Furutani-Horiuchi-Tamagaki (FHT) effective nucleon-nucleon force, which features central, spin-orbit, and tensor terms, with radial dependence described by Gaussian form factors~\cite{Furutani1979}. The Coulomb interaction between valence protons is treated exactly. The parameters of the GSM WS+FHT Hamiltonian are calibrated to reproduce the properties of very light nuclei, as detailed in Ref.~\cite{Jaganathen:2017}.  

The many-body basis for the Hamiltonian diagonalization is constructed from Slater determinants formed by Berggren basis states. Comprehensive details regarding the theoretical framework and numerical techniques required to solve the GSM eigenvalue problem are provided in the GSM reference book \cite{Michel:2021}.

\emph{Results and Discussion.}---
Table~\ref{tab:1} provides the experimental results of our final measurement of hyperfine splittings, achieved with an accuracy of approximately $10$ kHz. Our results showcase a four-fold increase in accuracy for the $2\,^3\!S_1$ state compared to prior experimental data, and a five to ten-fold increase in accuracy for the $2\,^3\!P_J$ state. As shown in the last row by the root-mean-square deviations $\bar\Delta$ between theory and experiment, these improved values align much more closely with the  theoretical calculations \cite{Qi2020} in the last column.

\begin{table}[htbp!]
  \caption{\label{tab:2} The determination of the Zemach radius contribution $\delta_{\mathrm{ZM}}$ by the hfs of the $2\,^3\!S_1$ state and the effective Zemach radius $\widetilde{R}_{\rm em}$, in fm.}
  \begin{ruledtabular}
  \begin{tabular}{lc}
  \multicolumn{1}{l}{Term}
  &\multicolumn{1}{c}{Value}\\
  \hline
  $A_{\rm the}$ (kHz)~\cite{Pachucki2023}       &\multicolumn{1}{r}{$7917508.1(1.3)$} \\
  $A_{\mathrm{exp}}$(kHz)                       &\multicolumn{1}{r}{$7926971.8(4.4)$} \\
  $a_e+\delta_{\rm QED}$~\cite{Pachucki2023}    &\multicolumn{1}{r}{$ 0.0015749(5) $} \\
  $\delta_{\mathrm{HO}}=A_{\mathrm{exp}}/A_{\rm the}-1$ &\multicolumn{1}{r}{$ 0.0011953(6) $} \\
  $\delta_{\mathrm{ZM}}$                        &\multicolumn{1}{r}{$-0.0003796(8) $} \\
  $\widetilde{R}_{\rm em}$ (This work)          &$3.35(1)       $ \\
  $\widetilde{R}_{\rm em}$~\cite{Pachucki2023}  &$3.33(3)       $ \\
  $\widetilde{R}_{\rm em}$~\cite{Qi2020}        &$3.38(3)       $ \\
  $\widetilde{R}_{\rm em}$~\cite{Puchalski2013} &$3.23(4)       $ \\
  \end{tabular}
  \end{ruledtabular}
\end{table}

We first consider the $2\,^3\!S_1$ state, where the contribution of the nuclear quadrupole moment vanishes. The hfs can thus be expressed in the form~\cite{Riis1994, Puchalski2013, Pachucki2023}
\begin{equation}\label{eq:4}
\begin{aligned}
E_{\rm hfs} (2\,^3\!S_1)= E_{F}(1+\delta_{\rm HO})\,,
\end{aligned}
\end{equation}
where $E_{F}=A \left\langle \bf{I} \cdot \bf{S} \right\rangle$ is the Fermi contact term~\cite{Qi2020}, $A$ is the hyperfine constant, $\bf{I}$ and $\bf{S}$ are the nuclear spin and the total electron spin. The higher-order correction $\delta_{\rm HO} = a_e+\delta_{\rm QED}+\delta_{\rm ZM}$, where $a_e$ is the anomalous magnetic moment of the electron, $\delta_{\rm QED}$ is a sum of higher-order QED corrections, and $\delta_{\rm ZM}$ is the contribution of the Zemach radius. For the Zemach radius, we use the effective radius $\widetilde{R}_{\rm em}$ recently introduced by Pachucki {\it et al.}~\cite{Pachucki2023} to account for other nuclear structure effects that would otherwise be omitted. It is defined by $\delta_{\rm ZM}=-2Z\widetilde{R}_{\rm em}$, where $Z$ is the nuclear charge.

Here, we use the method introduced by Pachucki {\it et al.}~\cite{Pachucki2023} to calculate the hyperfine constant $A$ for $2\,^3\!S_1$ state: $A = \frac{1}{6} \nu_{1/2-3/2} + \frac{3}{10} \nu_{3/2-5/2}$.
Combining the experimentally extracted value of $A_{\mathrm{exp}}$ with the theoretically calculated value of $A_{\rm the}$, we can determine the contribution of $\delta_{\rm HO}$. Furthermore, by separating the contributions of the anomalous magnetic moment and QED parts, we can obtain the $\delta_{\mathrm{ZM}}$ and determine the $\widetilde{R}_{\rm em}$, as shown in Table~\ref{tab:2}.
Our latest result for the effective Zemach radius agrees with previous values and shows a three-fold increase in accuracy.

\begin{figure*}[htbp!]
\includegraphics[scale=0.32]{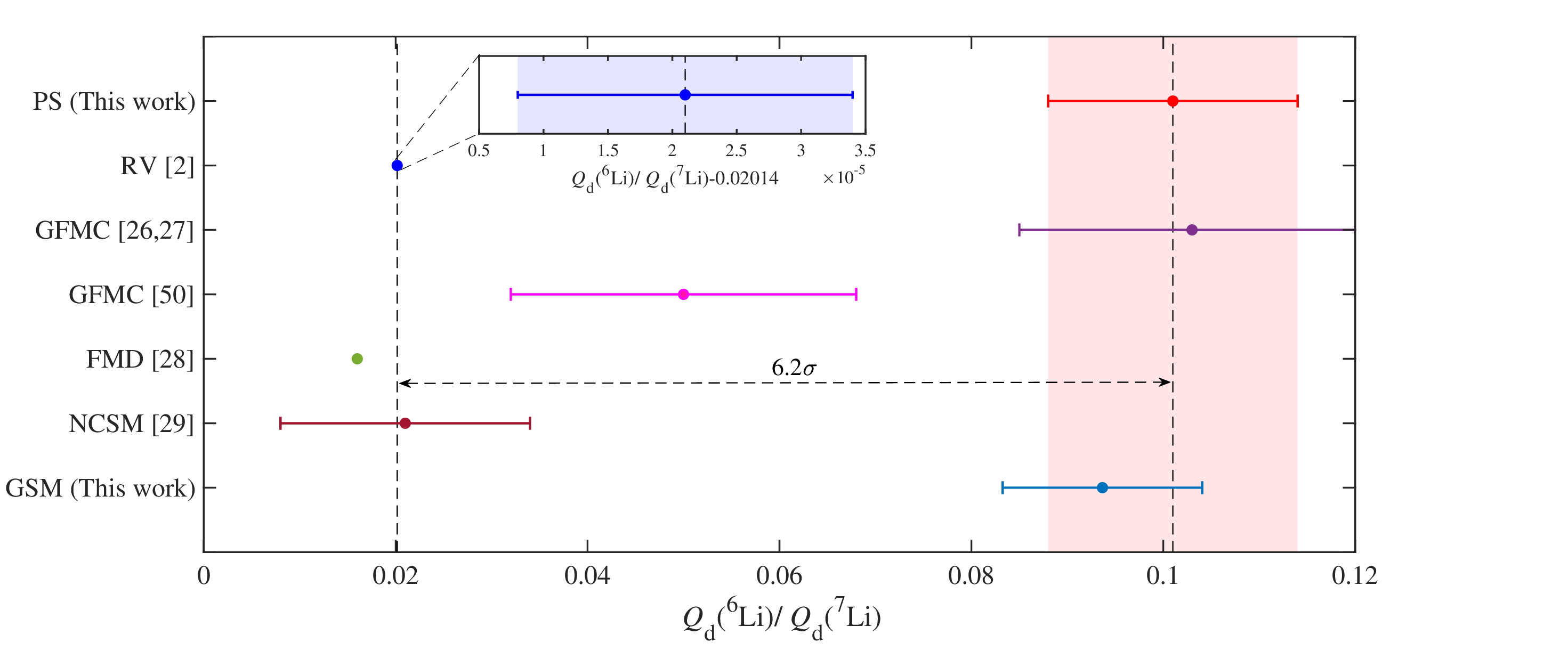}
\caption{The nuclear quadrupole moment ratio $Q_{d}(^6\text{Li})/Q_{d}(^7\text{Li})$. PS: Precision Spectroscopy; RV: Recommended Value (from molecular spectroscopy); GFMC: Green's Function Monte Carlo; FMD: Fermionic Molecular Dynamics; SM: Shell Model; NCSM: No-Core Shell Model; GSM: Gamow Shell Model.}
\label{fig:3}
\end{figure*}

Next, we employ a Taylor expansion to directly analyze the impact of the quadrupole moment $Q_{d}$ on the hfs of the $2\,^3\!P_J$ state~\cite{Qi2023}:
\begin{equation}\label{eq:8}
\begin{aligned}
\nu(Q_{d})=\nu(0)+XQ_{d}+YQ_{d}^2+O(Q_{d}^3)\;,
\end{aligned}
\end{equation}
where $\nu(0)$ is the hfs without $Q_{d}$, and its theoretical calculation has been discussed in detail in our previous work~\cite{Qi2020}. In the next term, $X$ is the linear expansion coefficient independent of $Q_{d}$ that can be evaluated through the first-order derivative of $\nu(Q_{d})$ at $Q_{d}=0$.
The quadratic term in the equation is used to evaluate the magnitude of higher-order contributions and determine whether retaining the linear term is sufficient at the current accuracy.

\begin{table}[htbp!]
\caption{\label{tab:3} 
Determination of the nuclear quadrupole moments $Q_d$ of $ ^6 \mathrm{Li} $ and $ ^7 \mathrm{Li} $ in $\mathrm{fm}^2$, with comparisons to results from various nuclear model calculations. PS: Precision Spectroscopy; Fit $B_2$: Fit the hyperfine $B_2$ constant; RV: Recommended Value (from molecular spectroscopy); ES: Electron Scattering; GFMC: Green's Function Monte Carlo; FMD: Fermionic Molecular Dynamics; SM: Shell Model; NCSM: No-Core Shell Model; GSM: Gamow Shell Model.}
\begin{ruledtabular}
\begin{tabular}{lcc}
\multicolumn{1}{l}{Method}
&\multicolumn{1}{c}{$Q_{d}(^6{\rm Li})$}
&\multicolumn{1}{c}{$Q_{d}(^7{\rm Li})$}\\
\hline
PS (This work)                      &$-0.39(5)  $  &$-3.86(5) $ \\
Fit $B_2$ (This work)~\cite{SM}     &              &$-3.84(7) $ \\
RV~\cite{Stone2016}                 &$-0.0806(6)$  &$-4.00(3) $ \\
ES~\cite{Voelk1991}                 &              &$-4.00(6) $ \\
GFMC~\cite{Pieper2001,Pieper2002}   &$-0.35(6)  $  &$-3.4(1)  $ \\
GFMC~\cite{Pastore2013}             &$-0.20(6)  $  &$-4.0(1)  $ \\
FMD~\cite{Nortershauser2011}        &$-0.061    $  &$-3.93    $ \\
SM~\cite{Radhi2014}                 &              &$-3.91    $ \\
NCSM~\cite{Forssen2009}             &$-0.066(40)$  &$-3.20(22)$ \\
GSM (This work)                     &$-0.385(26)$  &$-4.11(36)$ \\
\end{tabular}
\end{ruledtabular}
\end{table}

Our precision spectroscopy determinations of the nuclear quadrupole moments of $^6$Li and $^7$Li are summarized in Table~\ref{tab:3}, alongside comparisons with results from selected nuclear model calculations. Additional details on our determinations are provided in the Supplemental Material~\cite{SM}. The final values for the quadrupole moments $Q_d$ are $-0.39(5)$~$\mathrm{fm}^2$ for $^6$Li and $-3.86(5)$~$\mathrm{fm}^2$ for $^7$Li.  
Interestingly, our result for $^6$Li shows a significant deviation from the recommended value of $ -0.0806(6)$~$\mathrm{fm}^2$~\cite{Cederberg1998,Stone2016}, which was obtained using molecular spectroscopic methods to measure the ratio $Q_d(^6\mathrm{Li}) / Q_d(^7\mathrm{Li})$. 
However, our result is closer to the theoretical prediction of $-0.35(6) \, \mathrm{fm}^2 $ from GFMC calculations by Pieper \textit{et al.}~\cite{Pieper2001,Pieper2002}, while differing from the values of $ -0.061 \, \mathrm{fm}^2 $ reported in Ref.~\cite{Nortershauser2011} and $ -0.066(40) \, \mathrm{fm}^2$ in Ref.~\cite{Forssen2009}.  
For $^7$Li, our determined $Q_d$ value deviates from the recommended value by $2.3\sigma$. Nonetheless, it is consistent with results obtained using the shell model~\cite{Radhi2014}. 
Additionally, we applied the method of Pachucki \textit{et al.}~\cite{Puchalski2021} to fit the hyperfine constant $B_2$ and extract the nuclear quadrupole moment of $^7$Li, yielding $-3.84(7) \, \mathrm{fm}^2$~\cite{SM}, which is in agreement with our final determined value.  

Table~\ref{tab:3} also includes the results of our nuclear structure calculations using the GSM, where the effective charges for neutrons and protons are $0.25(5)$ and $1.25(5)$, respectively. The rationale behind these choices is provided in the Supplementary Material~\cite{SM}.  
Our theoretical results show excellent agreement with the values obtained from precision spectroscopy, particularly for $^6$Li, underscoring their consistency. However, a notable discrepancy is observed between the GSM and NCSM results, with the latter displaying a pronounced tendency toward smaller values. This highlights the importance of high-precision spectroscopic measurements as sensitive tests for nuclear structure theories, offering deeper insights into the underlying interactions within the nucleus.

Furthermore, we have calculated the ratio $Q_{d}(^6\text{Li})/Q_{d}(^7\text{Li}) = 0.101(13)$, which reveals a significant deviation from the value of $0.020161(13)$ derived from the molecular hyperfine spectra of $^6$Li$^{19}$F and $^7$Li$^{19}$F~\cite{Cederberg1998}, as depicted in Figure~\ref{fig:3}. This discrepancy represents a new challenge for the field of molecular spectroscopy and is expected to motivate further experimental measurements and theoretical investigations. The variations observed in the ratios from different nuclear theoretical approaches are particularly striking, as they are concentrated around two distinct and well-separated results. This warrants a deeper exploration within the context of nuclear structure theory.

\emph{Conclusion.}---
We measured the hyperfine splittings of $^7$Li at the $10$ kHz level and combined these results with our previous measurements for $^6$Li to extract nuclear structure information for both isotopes, including the Zemach radii and electric quadrupole moments. We have improved the accuracy of the Zemach radius of $^7$Li by a factor of three. Additionally, we successfully extracted the nuclear quadrupole moments of $^7$Li and $^6$Li, noting significant discrepancies from the current recommended values, particularly for $^6$Li, which shows a deviation of more than $6\sigma$. Further calculations using the GSM nuclear structure theory supported our findings, demonstrating good consistency for $^6$Li. In comparison with results from other theoretical models, we observed discrepancies among the predictions of different nuclear structure theories.
Finally, we computed the ratio $Q_d(^6\text{Li})/Q_{d}(^7\text{Li})$, which significantly differs from the LiF molecular spectroscopic results. These discrepancies are perplexing, and we call for additional experimental measurements in atomic and molecular spectroscopy, electron scattering, and related theoretical calculations pertaining to lithium.

Given the inherent challenges in determining nuclear quadrupole moments through nuclear physics theory—often relying on model-based calculations with significant uncertainties—the model-independent quadrupole moments derived from atomic spectra, as demonstrated in this work, are of critical importance. They provide a valuable benchmark for testing contemporary nuclear structure theories and may offer crucial insights into the exploration of the cosmological lithium problem.

The authors thank Qunfeng Chen, Yanqi Xu, and Huanyao Sun for technical support in optical frequency comb and electronic circuits.
We also appreciate Liyan Tang and Fangfei Wu for their valuable discussions on atomic structure theory, as well as Hankui Wang, Xiaofei Yang, and Youbao Wang for their helpful insights regarding nuclear physics theory.
This work is supported jointly by the National Natural Science Foundation of China (Grants No. 11934014, No. 92265206, No. 12393823, No. 12393821, No. 12121004, No. 12274423, No. 12204412, No. 12175199, No. 11974382, and No. 12174400),
CAS Project for Young Scientists in Basic Research (Grant No. YSBR-085 and No. YSBR-055),
Natural Science Foundation of Hubei Province (Grant No. 2022CFA013),
Science Foundation of Zhejiang Sci-Tech University (Grant No. 21062349-Y),
CAS Youth Innovation Promotion Association (Grants No. Y2022099).
Z. C. Y. and G. W. F. D. acknowledge research support by the Natural Sciences and Engineering Research Council of Canada.

$^*$These authors contributed equally to this work.

$^{\dag}$Email Address: tyshi@wipm.ac.cn

$^{\ddag}$Email Address: klgao@wipm.ac.cn


%

\end{document}